\documentstyle[12pt]{article}
\def\nabstar#1{\nabla\kern-0.5pt\smash{\raise 4.5pt\hbox{$\ast$}}
               \kern-4.5pt_{#1}}

\def\drvstar#1{\partial\kern-0.5pt\smash{\raise 4.5pt\hbox{$\ast$}}
               \kern-5.0pt_{#1}}

\def\newline{\relax\ifhmode\null\hfil\break\else\nonhmodeerr@\newline\fi}
\def\frac#1#2{{#1\over#2}}
\def\text#1{{\hbox{\rm #1}}}

\newcommand{\beq}{\begin{equation}}
\newcommand{\eeq}{\end{equation}}
\newcommand{\bea}{\begin{eqnarray}}
\newcommand{\eea}{\end{eqnarray}}

\def\BE{\begin{equation}}
\def\EE{\end{equation}}
\def\BA{\begin{eqnarray}}
\def\EA{\end{eqnarray}}
\def\BAN{\begin{eqnarray*}}
\def\EAN{\end{eqnarray*}}

\def\gm5{\gamma_5}

\input epsf.sty
\newdimen\psfigsize
\def\psfigure#1 #2 #3 #4 #5{
    \begin{figure}[tbh]
      \begin{center}
      \vbox{
        \null\vskip-0.2in\hskip#2
        \epsfxsize=#1
        \epsfbox{#4}
        \vskip -0.3in
        \caption {#5 \label{#3}}
        \vskip 0.0 true in plus 0.3 true in
      }
      \end{center}
   \end{figure}
}
\begin{document}
\thispagestyle{empty}
\begin{flushright}
NTUTH-02-505B \\
May 2002
\end{flushright}
\bigskip\bigskip\bigskip
\vskip 2.5truecm
\begin{center}
{\LARGE {Light quark masses in quenched QCD with exact chiral symmetry}}
\end{center}
\vskip 1.0truecm
\centerline{Ting-Wai Chiu and Tung-Han Hsieh}
\vskip5mm
\centerline{Department of Physics, National Taiwan University}
\centerline{Taipei, Taiwan 106, Taiwan.}
\centerline{\it E-mail : twchiu@phys.ntu.edu.tw}
\vskip 1cm
\bigskip \nopagebreak \begin{abstract}
\noindent

The parameters in the pseudoscalar meson mass formula
in quenched chiral perturbation theory to one-loop order
are determined by quenched lattice QCD with overlap Dirac operator,
and from which the light quark masses are determined
with the experimental inputs of pion and kaon masses,
and the pion decay constant. Our results are
$ m_{u,d} = 5.3 \pm 0.3 $ MeV, and $ m_s = 115 \pm 8 $ MeV,
in the $ \overline{\mbox{MS}} $ scheme at scale $ \mu = 2 $ GeV.

\vskip 1cm
\noindent PACS numbers: 11.15.Ha, 11.30.Rd, 12.38.Gc

\noindent Keywords : Chiral Perturbation Theory, Chiral Symmetry,
Lattice QCD, Overlap Dirac Quark.

\end{abstract}
\vskip 1.5cm

\newpage\setcounter{page}1



In the standard model, the quark masses are fundamental parameters which
have to be determined from high energy experiments. However, they cannot be
measured directly since quarks are confined inside hardrons, unlike
an isolated electron whose mass and charge both can be measured directly
from its responses in electric and magnetic fields.
Therefore, the quark masses can only be determined by comparing
a theoretical calculation of physical observables ( e.g., hadron masses )
with the experimental values. Evidently, for any field theoretic
calculation, the quark masses depend on the regularization,
as well as the renormalization scheme and scale.

One of the objectives of lattice QCD is to compute the hadron masses
nonperturbatively from the first principle, and from which
the quark masses are determined.
However, the performance of the present generation of computers is still
quite remote from what is required for computing the light hadron masses
at the physical ( e.g., $ m_\pi \simeq 140 $ MeV ) scale, on a lattice
with enough sites in each direction such that the discretization
errors as well as the finite volume effects are
both negligible comparing to the statistical ones.
Nevertheless, even with lattices of moderate sizes, lattice QCD
can determine the values of the parameters in the hadron mass
formulas of the ( quenched ) chiral perturbation theory.
Then one can use these formulas to evaluate the hadron masses at the
physical scale, as well as to determine the quark masses.

In quenched chiral perturbation theory \cite{Sharpe:1992ft,Bernard:1992mk},
the pion and kaon masses to one-loop order read as
\bea
\label{eq:mpi2}
m_\pi^2 &=& C m \left\{ 1 - \delta \left[
  \mbox{ln} \left( \frac{ C m}{\Lambda_{\chi}^2} \right) + 1 \right] \right\}
      + B m^2  \\
\label{eq:mk2}
\frac{m_K^2}{m_\pi^2} &=& \frac{m+m_s}{2m}
\left\{1+\delta\left[1-\frac{m}{m_s-m}\mbox{ln} \left( \frac{m_s}{m} \right)
\right] \right\}
\eea
where $ m $ denotes $ u $ and $ d $ quark bare masses in the isospin
limit ( $ m_u = m_d \equiv m $ ), $ m_s $ the $ s $ quark bare mass,
$ \Lambda_{\chi} = 2 \sqrt{2} \pi f_{\pi} $
( $ f_{\pi} \simeq 132 $ Mev ) the chiral cutoff,
$ \delta $ the coefficient of the quenched chiral logarithm, and
$ C $ and $ B $ are parameters. Similar to the Gell-Mann-Oakes-Renner relation,
\bea
\label{eq:GMOR}
m_\pi^2 = \frac{ 4 m \left< \bar\psi \psi \right> }{f_\pi^2} \ ,
\eea
the condensates to one-loop order are
\bea
\label{eq:cc_u}
\left< \bar{u} u \right> &=& \frac{ f_\pi^2 \ C } {4}
  \left\{ 1 - \delta \left[
  \mbox{ln} \left( \frac{ C m}{\Lambda_{\chi}^2} \right) + 1 \right]
      + \frac{B m }{C}  \right\}   \\
\label{eq:cc_s}
\frac{ \left< \bar{s} s \right> }{ \left< \bar{u} u \right> }
&=& \frac{m}{m_s} \left( \frac{ 2 m_K^2 }{ m_\pi^2 } - 1 \right)
\eea
where the parameter $ C $ is related to the $ u $ quark condensate
at the zeroth order
\bea
C = \frac{ 4 \left< \bar{u} u \right>_0 }{f_\pi^2} \ .
\eea
Note that the condensate $ \left< \bar{u} u \right> $
diverges logarithmically as the bare quark mass $ m \to 0 $,
a pathology of quenched approximation.
Thus the chiral symmetry breaking in QCD can only be properly addressed
with dynamical quarks.

With experimental values of pion and kaon masses as inputs to
(\ref{eq:mpi2}) and (\ref{eq:mk2}), one can determine the ratio of
light quark bare masses $ m_s / m $, but the absolute scale cannot
be fixed by quenched chiral perturbation theory. At this point,
lattice QCD plays the important role to fix the values of parameters
$ C $, $ B $ and $ \delta $, by measuring the pion mass
versus the bare quark mass.
Then the light quark masses $ m $ and $ m_s $
can be determined with experimental inputs of $ m_\pi $,
$ f_\pi $ ( to fix the lattice spacing $ a $ ), and $ m_K $.

Recently, we have determined the parameters $ C $, $ B $ and $ \delta $
\cite{Chiu:2002xm} in lattice QCD with
overlap Dirac operator \cite{Neuberger:1998fp,Narayanan:1995gw}.
For 100 gauge configurations generated with the Wilson
gauge action at $ \beta = 5.8 $ on the $ 8^3 \times 24 $ lattice, we
computed quenched quark propagators for 12 bare quark masses.
The pion decay constant [ $ f_\pi a = 0.0984(3) $ ] is extracted from
the pion propagator, and from which the lattice spacing is determined
to be
\bea
\label{eq:a}
 a  = 0.147(1) \ \mbox{fm}
\eea
By fitting (\ref{eq:mpi2}) to our data of $ ( m_\pi a )^2 $,
we obtained
\bea
\label{eq:delta_pi}
\delta &=& 0.2034(140) \ , \\
C a &=& 1.1932(182) \ ,  \\
\label{eq:B}
B &=& 1.1518(556)  \ .
\eea

Now inserting (\ref{eq:delta_pi}) and experimental values of
meson masses ( $ m_K = 495 $ MeV and $ m_\pi = 135 $ MeV )
into Eq. (\ref{eq:mk2}), we obtain the quark mass ratio
\bea
\label{eq:ms_m}
\frac{m_s}{m} = 21.92(23) \ ,
\eea
comparing with the ratio at the zeroth order ( $ \delta = 0 $ )
\bea
\label{eq:ms_m_0}
\left( \frac{m_s}{m} \right)_0 = 25.88
\eea

From (\ref{eq:mpi2}), with values of $ a $, $ \delta $,
$ Ca $ and $ B $ in (\ref{eq:a})-(\ref{eq:B}),
and $ m_\pi = 135 $ MeV, we obtain
\bea
\label{eq:m}
m = 6.3 \pm 0.3 \ \mbox{MeV} \ ,
\eea
which is inserted into (\ref{eq:ms_m}) to yield
\bea
\label{eq:ms}
m_s = 138 \pm 8 \ \mbox{MeV} \ .
\eea
Here we do not intend to estimate the finite volume and discretization
errors, in view of the exploratory nature of our present investigation
which will be followed by a systematic study with larger volumes, smaller
lattice spacings, and better statistics.

In order to transcribe above results (\ref{eq:m})-(\ref{eq:ms})
to their corresponding values in the usual renormalization scheme
$ \overline{\mbox{MS}} $ in high energy
phenomenology, one needs to compute the lattice renormalization
constant $ Z_m = Z_s^{-1} $, where $ Z_s $ is the renormalization
constant for $ \bar{\psi} \psi $. In general, $ Z_m $ should be
determined non-perturbatively. In this paper, we use the one loop
perturbation formula \cite{Alexandrou:2000kj} ( $ m_0 = 1.30 $ )
\bea
\label{eq:Zs}
Z_s = 1 + \frac{ g^2 }{ 4 \pi^2 }
\left[ \mbox{ln} ( a^2 \mu^2 ) + 6.7722 \right]
\eea
to obtain an estimate of $ Z_s $.
At $ \beta = 5.8 $, $ a = 0.147(1) $ fm, and $ \mu = 2 $ GeV, (\ref{eq:Zs})
gives $ Z_s = 1.19(1) $. Thus the quark masses in (\ref{eq:m})-(\ref{eq:ms})
are transcribed to
\bea
\label{eq:mMS}
m_{u,d}^{\overline{\mbox{MS}}}( 2 \mbox{ GeV} ) &=& 5.3 \pm 0.3 \mbox{ MeV }  \\
\label{eq:msMS}
m_s^{\overline{\mbox{MS}}}( 2 \mbox{ GeV} ) &=& 115 \pm 8 \mbox{ MeV}
\eea
It is expected that above values may become smaller if non-perturbative
renormalization is performed, however, the decreases should be less
than $ 10\% $. Nevertheless, the light quark masses in
(\ref{eq:mMS})-(\ref{eq:msMS}) are in good agreement with
the current lattice world average \cite{qmass_rev}.
We also note that in previous determinations
\cite{Giusti:2001pk,Hernandez:2001hq} of quark masses in
quenched QCD with overlap Dirac operator, $ ( m_{u,d} + m_s ) $ is
obtained \cite{Giusti:2001pk,Hernandez:2001hq} with
(\ref{eq:mpi2}) and (\ref{eq:mk2}) at the zeroth order
( $ \delta = 0 $, $ B = 0 $ ); while in Ref. \cite{Dong:2000mr},
$ m_{u,d} $ is obtained via the axial Ward identity.

In summary, the parameters of hadron mass formulas in ( quenched ) chiral
perturbation theory can be fixed by lattice QCD with overlap Dirac quarks.
This provides a viable approach to the determination of ( light ) quark
masses, as well as ( light ) hadron masses.




This work was supported in part by the National Science Council,
ROC, under the grant number NSC90-2112-M002-021,
and also in part by NCTS.


\vfill\eject

\end{document}